\documentclass[aps,prd,nofootinbib,reprint,superscriptaddress,longbibliography]{revtex4-1}
\usepackage{etoolbox}
\usepackage{dcolumn}% Align table columns on decimal point
\usepackage{bm}% bold math
\usepackage{graphics}
\usepackage{xcolor}
\usepackage{dirtytalk}
\usepackage{graphicx} %Include figure filesusepackage{graphicx} %Include figure files
\usepackage{blindtext}
\usepackage{amsmath}
\usepackage{amsfonts}
\usepackage{amssymb}
\usepackage{mathtools,halloweenmath}
\usepackage[%
colorlinks=true,
urlcolor=blue,
linkcolor=red,
citecolor=blue
]{hyperref}
\usepackage{physics}
\NewDocumentCommand{\tens}{t_}
{%
	\IfBooleanTF{#1}
	{\tensop}
	{\otimes}%
}
\NewDocumentCommand{\tensop}{m}
{%
	\mathbin{\mathop{\otimes}\displaylimits_{#1}}%
}
\usepackage{natbib}
\begin{document}
\title{\bf Effect of non-conformal deformation on the gapped quasi-normal modes and the holographic implications}
\vskip 1cm
\author{Ashis Saha}
\email{ashis.hepth@gmail.com}
\affiliation{Department of Physics, School of Basic and Applied Sciences,\\
	Adamas University, Barasat, Kolkata 700126, India}
%\affiliation{Physics and Applied Mathematics Unit,\linebreak
%	Indian Statistical Institute,\linebreak
%	 203 B.T. Road, Kolkata 700108, India}
\author{Sunandan Gangopadhyay}
\email{sunandan.gangopadhyay@bose.res.in}
\affiliation{Department of Astrophysics and High Energy Physics,\linebreak
	S.N.~Bose National Centre for Basic Sciences,\linebreak
	 JD Block, Sector-III, Salt Lake, Kolkata 700106, India}	

	\begin{abstract}
		\noindent The spectral curve of quasinormal modes for a massive real scalar field in the background of a non-conformal black brane geometry has been obtained by utilizing a Frobenius type near-horizon expansion. The gauge/gravity duality maps this to the computation of spectral curve of a massive scalar operator $\mathcal{O}_{\phi}$ for a large-$N$ conformal field theory with irrelevant type non-conformal deformation. In this context, non-conformality has been holographically introduced by using the Einstein-dilaton theory with Liouville type dilaton potential as the bulk theory. It has been observed that the obtained quasinormal modes are characterized by specific gapped dispersion relations. The pole-skipping points have also been computed and classified based upon different dispersion relations satisfied by them. The effect of non-conformality is evident from these results. The radius of convergence of the derivative expansion in the momentum space is then computed from the critical points of the spectral curve. It has been observed that presence of non-conformality increases the domain of applicability of the derivative expansion in momentum space, as it increases the radius of convergence for a given conformal dimension. The comparison between the convergence radii and the absolute momenta corresponding to lowest order pole-skipping points also leads to some interesting findings.  
	\end{abstract}
\maketitle
\section{Introduction}
\noindent The study of relativistic hydrodynamics (RH) has always been a matter of immense interest due to its wide applicability, ranging from quark-gluon plasma (QGP) created at the Relativistic Heavy Ion Collider (RHIC), Large Hadron Collider (LHC) \cite{Heinz:2013th,Noronha-Hostler:2015wft,Florkowski:2017olj}, to the interior structure of neutron stars \cite{Faber:2001nt,Shibata:2017jyf,vanEysden:2018kbc}. Fundamentally, relativistic hydrodynamics provides a framework to understand the collective macroscopic behaviour at the large-scale, long-wavelength limit (small frequency and small momenta), in terms of the conserved quantities, such as, energy, momentum and charge \cite{Romatschke:2017ejr}. However, a direct derivation of the hydrodynamic formulation from the postulates of quantum physics is still a challenging task \cite{Kovtun:2012rj}. The core idea of RH relies upon the conservation laws $\grad_{\alpha}T^{\alpha\beta}=0$, $\grad_{\alpha}J^{\alpha}=0$, and the constitutive relations which are simply the rules that tell us how the macroscopic currents ($T^{\alpha\beta},J^{\alpha}$) depend on the local hydrodynamic variables (four-velocity of the fluid $u^{\mu}$, temperature of the fluid $T$, chemical potential $\mu$, etc.). The constitutive relations can be understood from the following general asymptotic expansions
\begin{eqnarray}\label{series1}
T^{\alpha\beta}&=&T^{\alpha\beta}_{(0)}+T^{\alpha\beta}_{(1)}+T^{\alpha\beta}_{(2)}+..\nonumber\\
J^{\alpha}&=&J^{\alpha}_{(0)}+J^{\alpha}_{(1)}+J^{\alpha}_{(2)}+...~.
\end{eqnarray}
In the above, the numbers in the subscript denote the order of RH which is defined from the fact that how many derivatives (gradient) of the macroscopic variable are being considered. In simple sense, if we restrict ourselves only to the zeroth order, then we have $T^{\alpha\beta}\approx T^{\alpha\beta}_{(0)}$ and $J^{\alpha}\approx J^{\alpha}_{(0)}$, where $T^{\alpha\beta}_{(0)}$ and $J^{\alpha}_{(0)}$ are proportional to the hydrodynamic variables and not their derivative (gradient). This in turn gives the theory for relativistic perfect fluid governed by the Euler equations
\begin{eqnarray}
T^{\alpha\beta}\approx T^{\alpha\beta}_{(0)} &=& \left(\epsilon+p\right) u^{\alpha}u^{\beta}+pg^{\alpha\beta}\nonumber\\
J^{\alpha}\approx J^{\alpha}_{(0)}&=&\mathcal{N}u^{\alpha}
\end{eqnarray}
where $\epsilon, p$ and $\mathcal{N}$ represent (local) energy density, pressure and charge density respectively. On the other hand, if we consider terms upto first order ($\sim \mathcal{O}(T^{\alpha\beta}_{(1)},J^{\alpha}_{(1)})$) in the series given in eq.\eqref{series1}, we get the relativistic Navier-Stokes equations and the terms $T^{\alpha\beta}_{(1)}$ and $J^{\alpha}_{(1)}$ contains first derivative (gradient) of the hydrodynamic variables which include the viscous effects of the fluid. This reads
\begin{eqnarray}
T^{\alpha\beta}_{(1)}&=&-\eta \left[2\Delta^{\alpha\rho}\Delta^{\beta\delta}\grad _{<\rho}u_{\delta >}\right]-\zeta\Delta^{\alpha\beta}\left(\div u\right)\nonumber\\
J^{\alpha}_{(1)} &=& -\kappa T \Delta^{\alpha\rho}\grad_{\rho}\left(\frac{\mu}{T}\right)~.
\end{eqnarray}
In the above expressions, the transport coefficients $\eta$, $\xi$ and $\kappa$ are shear viscosity, bulk viscosity and conductivity respectively. Similarly, at the second order, we obtain the M\"uller-Israel-Stewart (MIS) theory \cite{1967ZPhy..198..329M,Israel:1976efz,Israel:1976tn,Israel:1979wp} of relativistic fluid which resolves the problem of acausality at the first order \cite{Hoult:2020eho,Kovtun:2019hdm,Bemfica:2019knx,Bemfica:2020zjp}. These observations suggest that the hydrodynamic derivative (gradient) expansion provides an effective macroscopic description for relativistic fluids, and the choice of the expansion order is left to our discretion. Furthermore, the coefficients
of this expansion are the so called transport coefficients which capture the information of the underlying microscopic theory. However, it is to be mentioned that the gradient expansion is typically asymptotic rather than convergent. This means that while including higher-order terms can improve accuracy up to an optimal order, beyond this point, terms grow and the series diverges (in real space). This in turn means it has a zero radius of convergence. However, by using Bor\'el transformations one obtains a series with a finite radius of convergence set by the lightest quasi-normal‑mode frequency. Finally, a suitable Bor\'el integral ``resums" the series into a single, well‑defined function of time, reproducing both the familiar viscous corrections and the exponentially suppressed transient physics and thereby revealing the unique hydrodynamic attractor toward which the system evolves. \cite{Heller:2013fn,Basar:2015ava,Heller:2015dha,Aniceto:2015mto,Heller:2016rtz,Aniceto:2018uik}. By mapping the hydrodynamic gradient expansion in momentum space it can be shown that the radius of convergence emerges from the analytical structure of the retarted Green's function. To be precise, within kinetic theory in the relaxation time approximation, the retarded Green's function has both (hydrodynamic) poles and branch points and collision of these points (for sound channel: collision of hydrodynamic pole-branch point and for sound channel: pole-pole collision) gives us the radius of convergence for the hydrodynamic gradient expansion in momentum space \cite{Romatschke:2015gic,Heller:2020hnq}. Due to the strongly coupled nature of QGP \cite{Heinz:2004ar,Romatschke:2007mq,Heinz:2013th}, use of a non-perturbative approach in context of RH, such as gauge/gravity duality \cite{Maldacena:1997re,Witten:1998qj,Aharony:1999ti,Son:2002sd} has been very fruitful so far \cite{Policastro:2001yc,Policastro:2002se,Policastro:2002tn,Kovtun:2004de,Baier:2007ix,Heller:2007qt,PhysRevD.77.046006}.\\
This strong/weak duality has been helping us to understand various properties of a strongly coupled quantum gauge theory in terms of a weakly coupled classical gravitational dual at one dimension higher. Motivated by this ``holographic" set up, in \cite{Withers:2018srf,Grozdanov:2019kge,Grozdanov:2019uhi}, it has been shown that for gauge theories with gravitational dual, the radius of convergence of the hydrodynamic derivative expansion (in complex momentum-complex frequency space) can be obtained from the critical points of the spectral curve corresponding to the quasi-normal modes (QNMs). This observation was further extended in various different holographic scenarios \cite{Jansen:2020hfd,Abbasi:2020ykq,Abbasi:2021fcz,Baggioli:2020loj,Abbasi:2020xli,Grozdanov:2021jfw,Asadi:2021hds,Heydari:2024cun}. In the context of gauge/gravity duality, the QNMs are holographically dual to the poles of the retarted Green's function of the relevant operator (dual to the perturbation field considered in the bulk). Physically, these poles govern how small perturbations of the dual thermal QFT state decay back to equilibrium. Some insightful works regarding (gravitational) QNMs and hydrodynamics can be found in \cite{PhysRevD.73.064034,Miranda:2008vb,Morgan_2009,Diles_2020,diles2023thirdorderrelativisticfluiddynamics}. In general, the lowest‑lying QNMs are the most significant, as they correspond to the slowest relaxation channels, such as hydrodynamic modes. However, consideration of all the QNMs give us the associated spectral curve (under the correct boundary condition) $\mathcal{F}(\omega,k^2)=0$. Furthermore, by solving the spectral curve in the hydrodynamic limit (large wavelength limit) one obtains the hydrodynamic dispersion relations (gapless: $\omega\rightarrow0$ as one considers $k\rightarrow0$) such as (sound channel) $\omega(k)=\pm v_s k-\frac{i}{2}\Gamma_s k^2+...$ and (shear diffusion channel) $\omega(k)=-iDk^2+...$~. It is to be mentioned that in order to obtain these gapless hydrodynamic dispersion relations holographically, one has to consider gravitational perturbation in the bulk.\\
However, apart from these gapless sound or shear hydrodynamic modes, physical systems also has gapped modes ($\omega\neq0$ for $k\rightarrow0$) \cite{Kovtun:2005ev,Berti:2009kk,Festuccia:2008zx}. These gapped modes play a crucial role. The reasons are as follows. Although hydrodynamics governs the late‐time, long‐wavelength dynamics in terms of the gapless sound and diffusion modes, it is now well established that this hydrodynamic sector does not exhaust the full relaxation spectrum of the field theories \cite{Baggioli:2019jcm}. Further, the expansion in gradients only converges up to the point where a non‑hydrodynamic (gapped) mode collides with a hydrodynamic pole in the complex‑momentum plane. Physically, these gapped modes set the fast relaxation scales which is similar to the transient modes in a QGP just after a heavy‑ion collision and appear as metastable resonances (such as, holographic ``mesons" or glueballs in bottom‑up QCD models \cite{Janik:2005zt,Bigazzi:2011fx}, or pinned phonon resonances in holographic condensed‑matter systems \cite{Herzog:2009xv}). Tracking where and how these massive modes approach the hydrodynamic branch not only fixes the radius of convergence of the gradient series but also tells us why and when hydrodynamics ceases to be a valid effective description. On the other hand, in context of AdS/CFT correspondence, in addition to the familiar hydrodynamic poles, holographic retarded Green’s functions generically contain an infinite tower of non-hydrodynamic quasi-normal modes, including gapped modes that remain at finite complex frequency even in the zero-momentum limit. These modes encode fast, transient relaxation processes that control the approach to local equilibrium and set the intrinsic timescales beyond which hydrodynamics becomes applicable \cite{Baggioli:2019abx,Baggioli:2023tlc}. Crucially, the analytic structure associated with such gapped excitations governs the breakdown of the hydrodynamic derivative expansion, as the radius of convergence is fixed by collisions between hydrodynamic and non-hydrodynamic modes in the complex momentum plane. Following \cite{Abbasi:2020xli}, our aim in this work is to compute the radius of convergence of the hydrodynamic derivative expansion about the gapped modes. In context of study related to QNMs, another interesting phenomena which has gained a lot of attention recently is the pole-skipping phenomenon \cite{Blake:2018leo,Blake:2019otz,Natsuume:2019xcy,Natsuume:2019sfp,Natsuume:2019vcv,Jeong:2021zsv,Huh:2021ppg,Jeong:2021zhz,Ahn:2024gjh}. In the complex-$(\omega,k)$ plane, there exist some special points at which the holographic retarded Green's function becomes non-unique or ill-defined. Further, these points become more interesting when one considers the energy-density retarded Green's function, as the pole-skipping points of this holographic correlator are related to the parameters of chaos, such as Lyapunov exponent \cite{Maldacena:2015waa} and butterfly velocity \cite{Roberts:2016wdl}. This is a very important result as it enables us to study chaotic systems at the two-point level, apart from the traditional approach of computing the $4$-pt functions, such as, OTOC. In this work, we will try to probe the relation between the scalar-field pole-skipping points and the radius of convergence of the gradient expansion.\\
Another important thing which we will consider in this work is non-conformality. From the QGP phenomenology perspective, incorporating non-conformality is important because heavy‑ion collisions probe a fluid whose dynamics are governed by a running coupling, a pronounced trace anomaly, and a sharp departure from scale invariance around the crossover temperature. Further, for a conformal fluid one has zero bulk viscosity, however, in actual scenario where scale-invariance is absent, it shows a sharp peak around the cross-over temperature \cite{Karsch:2007jc}. In this work, we consider a particular black brane solution of Einstein-dilaton theory which has a Liouville type dilaton potential $V(\phi)$ \cite{Charmousis:2001nq}. Some interesting studies related to this model can be found in \cite{Kulkarni:2012re,Kulkarni:2012in,Park:2012lzs,Park:2015afa,Saha:2024bpt}. In context of gauge/gravity duality, it is a well-known fact that the asymptotic
limit of the black brane solution represents the UV fixed point of the dual gauge theory. On the other hand, the bulk diffeomorphisms are dual to the Weyl transformations of the boundary theory. Now inclusion of an irrelevant operator (deformation) in the dual gauge theory can break the conformality of the boundary
theory away from the UV fixed point and holographically this can lead us to a
non-AdS, warped geometry in the bulk. In short, Einstein-dilaton theory with
Liouville type dilaton potential probes non-conformal (or more precisely, irrelevant) deformation of large N boundary gauge theory \cite{Broccoli:2021icm}. To summarize, the novelty of this work lies in the understanding the behaviour of the gapped modes and how they are modified by non-conformal deformations so that one can assess the regime of validity of hydrodynamics in realistic, strongly coupled system that lack scale invariance. Our results clarify the dynamical role of non-hydrodynamic gapped modes in non-conformal holographic plasmas and establish a direct link between non-conformality, transient relaxation scales, and the analytic structure of holographic Green’s functions.\\
The organization of this paper is as follows. In section \eqref{Sec1},  we compute the spectral curve of the quasi-normal modes for a real, massive scalar field in the background of non-conformal black brane geometry. This also holographically leads us to the spectral curve of the dual scalar operator in presence of non-conformal deformation. We then proceed to compute the pole-skipping points and locate them in the complexified frequency-momenta plane. It has been given in section \eqref{Sec2}. In section \eqref{Sec4}, the radius of convergence of derivative expansion about the gapped quasi-normal modes is then computed by obtaining the critical points of the associated spectral curve. Furthermore, in section \eqref{Sec5}, we then compare these obtained values of convergence radii with the absolute momenta associated to the lowest order pole-skipping points and get some vital insights. We summarize our findings in section \eqref{Sec6}.
\section{(Scalar) quasi-normal modes and spectral curve of the dual operator}\label{Sec1}
\noindent The Einstein-Hilbert action corresponding to the ($d+1$)-dimensional Einstein-dilaton theory with Liouville type potential reads \cite{Kulkarni:2012re,Kulkarni:2012in,Park:2012lzs,Park:2015afa}
\begin{eqnarray}\label{bulkaction}
S_{\mathrm{EH}}=\frac{1}{16\pi G_N^{d+1}}\int d^{d+1}x \sqrt{-g}\left[R-2(\partial\Phi)^2-V(\Phi)\right]\nonumber\\
\end{eqnarray}
where $V(\Phi)=2\Lambda e^{\eta\Phi}$ is the Liouville type dilaton potential. Here $\Lambda<0$ represents the cosmological constant and $\eta$ denotes the non-conformal parameter which captures the deviation of the system from conformality. Further, $\eta<\sqrt{\frac{8d}{(d-1)}}$ \cite{Gubser:2000nd,Gouteraux:2011ce}. This upper bound is also denoted as the Gubser bound on $\eta$ for this Einstein-dilaton model and it was given from the perspective of thermodynamic stability of the black brane. In \cite{Saha:2024bpt}, the same upper bound on $\eta$ was given from the perspective of quantum chaos as at this partcular value the parameters of quantum chaos, namely, Lyapunov exponent, butterfly velocity and entanglement velocity vanishes. The corresponding Einstein field equations and the equation of motion for the dilaton field reads
\begin{eqnarray}\label{EOM}
R_{\mu\nu}-\frac{1}{2}g_{\mu\nu}R+\frac{1}{2}g_{\mu\nu}V(\Phi)&=&2\partial_{\mu}\Phi\partial_{\nu}\Phi-g_{\mu\nu}(\partial\Phi)^2\nonumber\\
\frac{1}{\sqrt{-g}}\partial_{\mu}\left(\sqrt{-g}g^{\mu\nu}\partial_{\nu}\Phi\right)&=&\frac{1}{4}\frac{\partial V(\Phi)}{\partial \Phi}~.
\end{eqnarray}
In order to solve these equations, one generally considers a logarithmic dilaton profile of the following form \cite{Kulkarni:2012re,Park:2012lzs}
\begin{eqnarray}
	\Phi(r)=\Phi_0-\Phi_1 \log(r)
\end{eqnarray}
where $\Phi_0$ and $\Phi_1$ are integration constants. As $\Phi_0$ can be absorbed into the cosmological constant, one can set $\Phi_0=0$ without any loss of generality. Now, proceeding with a planar ansatz for the metric and keeping in mind the logarithmic dilaton profile, one
obtains the following non-conformal black brane geometry (NCBB) (we have set $G_N=1$ and AdS radius $R=1$, for the sake of simplicity) \cite{Kulkarni:2012re,Kulkarni:2012in,Park:2012lzs,Park:2015afa,Saha:2024bpt}
\begin{eqnarray}\label{EdBB}
ds^2&=&-r^{2p}f(r) dt^2+\frac{dr^2}{r^{2p}f(r)}+r^{2p}\sum_{i=1}^{d-1}dx_{i}^2\\
f(r)&=&1-\left(\frac{r_+}{r}\right)^c\nonumber
\end{eqnarray}
where we have used
\begin{eqnarray}\label{NCparameter}
p=\frac{8}{8+(d-1)\eta^2};~c=\frac{8d-(d-1)\eta^2}{8+(d-1)\eta^2}~.
\end{eqnarray}
As mentioned earlier, the asymptotic limit of the above given black brane is not AdS, rather it is warped geometry and this in turn means that the boundary field theory is not conformal. Further, it can be observed that in the limit $\eta\rightarrow0$, one obtains the AdS$_{d+1}$-Schwarzschild black brane solution.The Hawking temperature of the black brane geometry (given in eq.\eqref{EdBB}) reads
\begin{eqnarray}\label{HawkingT}
T = \frac{k}{2\pi}= \left(\frac{c}{4\pi}\right)r_+^{2p-1}
\end{eqnarray}
where $k$ is the surface gravity. It is to be observed that for $\eta=\sqrt{\frac{8d}{(d-1)}}$, the Hawking temperature of the black brane is zero, irrespective of the value of $r_+$. For the simplification of the subsequent calculation, we now make a suitable coordinate transformation of the following form
\begin{eqnarray}
\left(\frac{r_+}{r}\right)^c=u^2;~dr=-\left(\frac{2}{c}\right)\frac{r_+}{u^{1+\frac{2}{c}}}~du~.
\end{eqnarray}
Under these transformations, the metric takes the following form
\begin{eqnarray}\label{NCBB2}
ds^2&=& \left(\frac{r_+}{u^{\frac{2}{c}}}\right)^{2p}\Big[-f(u)dt^2+dx_i dx^i\Big]\nonumber\\
&&+\left(\frac{2u^{\frac{2p}{c}-\frac{2}{c}-1}}{cr_+^{p-1}}\right)^{2}\frac{du^2}{f(u)}~.
\end{eqnarray}
It is to be mentioned that in the above form of the metric, the lapse function reads $f(u)=1-u^2$. We now perturb the boundary gauge theory by including a scalar operator $\mathcal{O}_{\phi}$. This can be holographically realised by perturbing the NCBB geometry with a minimally-coupled massive (real) scalar field $\phi$, which is dual to the massive scalar operator $\mathcal{O}_{\phi}$. The EOM (usual Klein-Gordon equation) for the scalar field reads
\begin{eqnarray}\label{KGeq}
\frac{1}{\sqrt{-g}}\partial_{\mu}(\sqrt{-g}\partial^{\mu}\phi)=m^2\phi~.
\end{eqnarray}
In the above equation $\sqrt{-g}=\left(\frac{2c}{u}\right)\left(\frac{r_+}{u^{\frac{2}{c}}}\right)^{1+p(d-1)}$ which can be easily derived from the metric given in eq.\eqref{NCBB2}.  To proceed further, we take the plane-wave ansatz for the bulk field $\phi(t,u,x^i)$, that is,
\begin{eqnarray}\label{planewave}
\phi(t,u,x^i)=\int d^dk e^{-i\omega t+ik_ix^i}\psi_{\omega,k}(u)~.
\end{eqnarray}
By using the above ansatz and metric tensor components from the metric given in eq.\eqref{NCBB2}, one obtains the following radial equation
	\begin{eqnarray}\label{RadEq1}
\mathcal{A}_1(u)\psi^{\prime\prime}_{\omega,k}(u)+\mathcal{A}_2\psi^{\prime}_{\omega,k}(u)+\mathcal{A}_3\psi_{\omega,k}(u)=0
\end{eqnarray}
	where,
\begin{eqnarray}
	\mathcal{A}_1(u)&=&\left(\frac{c}{2}r_+^{(p-1)}u^{1+\frac{2}{c}-\frac{2p}{c}}\right)^2f(u)\nonumber\\
	\mathcal{A}_2(u)&=&\left(\frac{c}{2}r_+^{(p-1)}\right)^2\Big[f^{\prime}(u)u^{2+\frac{4}{c}-\frac{4p}{c}}\nonumber\\
	&&+\left(1+\frac{2}{c}-\frac{2p}{c}\right)f(u)u^{1+\frac{4}{c}-\frac{4p}{c}}\Big]\nonumber\\
	\mathcal{A}_3(u)&=&\left(\frac{u^{\frac{2}{c}}}{r_+}\right)^{2p}\frac{1}{f(u)}\left[\omega^2-k^2f(u)-m^2f(u)\left(\frac{r_+^p}{u^{\frac{2p}{c}}}\right)^2\right]~.\nonumber\\
\end{eqnarray}	
Now, our aim here is to compute the entire spectrum of the QNMs, and to do that we follow the approach shown in \cite{Horowitz:1999jd,Berti:2009kk}. First, we rewrite the radial eq.\eqref{RadEq1} in a suitable form, which reads
\begin{eqnarray}\label{RadEq2}
(u-1)\mathcal{S}_{(u)}\psi^{\prime\prime}_{\tilde{\omega},\tilde{k}}+\mathcal{T}_{(u)}\psi^{\prime}_{\tilde{\omega},\tilde{k}}+\frac{\mathcal{P}_{(u)}}{(u-1)}\psi_{\tilde{\omega},\tilde{k}}=0~.
\end{eqnarray}
The new functions $\mathcal{S}_{(u)}$, $\mathcal{T}_{(u)}$ and $\mathcal{P}_{(u)}$ are of the following form
\begin{eqnarray}\label{expansion1}
\mathcal{S}_{(u)}&=&(u+1)^2u^{2+\frac{4}{c}-\frac{4p}{c}}\nonumber\\
\mathcal{T}_{(u)}&=&-(u+1)u^{1+\frac{4}{c}-\frac{4p}{c}}\left[uf^{\prime}(u)+\left(1+\frac{2}{c}-\frac{2p}{c}\right)f(u)\right]\nonumber\\
\mathcal{P}_{(u)}&=&u^{\frac{4p}{c}}\tilde{\omega}^2-\left(u^{\frac{4p}{c}}\tilde{k}^2+\left(\frac{2}{c}\right)^2\tilde{m}^2\right)f(u)~.
\end{eqnarray}	
In computing the above, we have introduced the following rescalings
\begin{eqnarray}
\tilde{\omega}=\frac{\omega}{2\pi T};~\tilde{k}=\frac{k}{2\pi T};~\tilde{m}=\frac{m}{r_+^{p-1}}~.
\end{eqnarray}
\noindent We now assume the following near-horizon expansions
\begin{eqnarray}\label{NHexpansion}
(i)~&&\left\{\mathcal{S}_{(u)},\mathcal{T}_{(u)},\mathcal{P}_{(u)}\right\}=\sum_{m=0}^{\infty}\left\{s_m,t_m,p_m\right\}(u-1)^m\nonumber\\
(ii)~&&\psi_{\tilde{\omega},\tilde{k}}(u)=(u-1)^{\alpha}\sum_{n=0}^{\infty}a_n(\tilde{\omega},\tilde{k}^2,\tilde{m})(u-1)^n
\end{eqnarray}
where $\alpha$ classifies the ingoing and outgoing solutions. This can be realised in the following way. By using the above expansions in eq.\eqref{RadEq2}, if we focus only at the leading behaviour ($n=0,m=0$), we get
 \begin{eqnarray}
 s_0\alpha (\alpha-1)&+&t_0\alpha+p_0=0\nonumber\\
 \implies\alpha&=&\pm \frac{i\tilde{\omega}}{2}
 \end{eqnarray}
 where we have used the fact $\left\{s_0,t_0,p_0\right\}\equiv\left\{\mathcal{S}_{(u)},\mathcal{T}_{(u)},\mathcal{P}_{(u)}\right\}\vert_{u=1}$. Here, we are only interested in the ingoing solution, so we choose $\alpha=-\frac{i\tilde{\omega}}{2}$. This is due to the fact that QNMs are poles of the retarded Green's function of the operator $\mathcal{O}_{\phi}$ which holographically corresponds to the ingoing solution. We now focus on the whole structure of the near-horizon and obtain the following equation
 \begin{widetext}
 \begin{eqnarray}\label{sum}
 &&\sum_{n=0}^{\infty}\sum_{m=0}^{\infty}a_n\left[s_m\left\{n(n-1)+2\alpha n+\alpha (\alpha-1)\right\} +t_m\left(\alpha+n\right)+p_m\right]\left(u-1\right)^{m+n}=0\nonumber\\
 \implies &&\sum_{j=0}^{\infty}\sum_{k=0}^{j}a_k\left[s_{j-k}\left\{k(k-1)+2k\alpha+\alpha (\alpha-1)\right\} +t_{j-k}\left(\alpha+k\right)+p_{j-k}\right]\left(u-1\right)^j=0\nonumber\\
 \implies &&a_j=-\frac{1}{\Lambda_{j,\alpha}}\sum_{k=0}^{j-1}a_k\left[s_{j-k}\left\{k(k-1)+2k\alpha+\alpha (\alpha-1)\right\} +t_{j-k}\left(\alpha+k\right)+p_{j-k}\right]
 \end{eqnarray}
where in the last line we have introduced $\Lambda_{j,\alpha}=s_0\left\{j(j-1)+2j\alpha+\alpha(\alpha-1)\right\}+t_0\left(\alpha+j\right)+p_0$. 	
 \end{widetext}
Now if we use the above expression of $a_j$ along with the Dirichlet boundary condition at $u=0$ in the ansatz taken for $\psi_{\tilde{\omega},\tilde{k}}$ (given in eq.\eqref{NHexpansion}), we get the spectral curve for the dual operator $\mathcal{O}_{\phi}$. This reads
\begin{eqnarray}\label{SCurve}
\mathcal{S}_{\phi}(\tilde{\omega},\tilde{k}^2,m):=\sum_{j=0}^{\infty}(-1)^ja_j(\tilde{\omega},\tilde{k}^2,m)=0
\end{eqnarray}
where the expression for $a_j$ is given in eq.\eqref{sum}. By considering a large enough but finite number terms, one can obtain the desired QNMs numerically \cite{Nunez:2003eq,Abbasi:2020xli}. Before proceeding, we briefly address the mass-dimension relation for the field $\phi$ and its dual operator $\mathcal{O}_{\phi}$.
\section{Asymptotic behaviour of the massive bulk field and the mass-dimension relation}
In order to do this, we first consider the asymptotic form of the black brane geometry given in eq.\eqref{EdBB}. This reads
\begin{eqnarray}
ds^2&=& r^{2p}\left(-dt^2+\sum_{i=1}^{d-1}dx_{i}^2\right)+\frac{dr^2}{r^{2p}}~.
\end{eqnarray}
The above metric once again confirms the non-AdS, warped asymptotic structure of the non-conformal black brane geometry. Further, we assume $\phi\sim \phi(r)$ (as we are only interested in the bulk-directional behaviour). This leads us to the following equation of motion
\begin{eqnarray}\label{MassDimEq}
r^{2p}\phi^{\prime\prime}(r)+p(d+1)r^{2p-1}\phi^{\prime}(r)-m^2\phi(r)=0~.
\end{eqnarray}
The above equation can be solved easily in the conformal limit ($\eta\rightarrow0$, which yields $p=1$) and it produces the known mass-dimension relation (assuming $\phi\sim \frac{1}{r^{\Delta_{\phi}}}$)
\begin{eqnarray}\label{MassDimStand}
\Delta_{\phi}=\left(\frac{d}{2}\right)\pm\sqrt{m^2+\frac{d^2}{4}}~.
\end{eqnarray}
However, as it can be seen it is not so trivial to solve for $\eta\neq0$. It is to be observed that for $m=0$, eq.\eqref{MassDimEq} simplifies to the following form
	\begin{eqnarray}
	r^{2p}\phi^{\prime\prime}(r)+p(d+1)r^{2p-1}\phi^{\prime}(r)=0~.
	\end{eqnarray}
	Surprisingly, the above equation has a scaling symmetry and one can proceed to solve it with the ansatz $\phi(r)\sim \frac{1}{r^{\Delta_{\phi}}}$. This leads to the following values for the dimension 
	\begin{eqnarray}
	\Delta_{\phi}=p(d+1)-1,~0~.
	\end{eqnarray}
	The above relation is interesting because, although we are dealing with a black brane geometry which has a non-AdS warped asymptotic structure, it is possible to derive a definite value for the dimension of the dual operator as long as the mass of the operator is zero. Furthermore, the derived result expresses $\Delta_{\phi}$ in terms of the non-conformal parameter $\eta$, given the dependence of $p$ on $\eta$ shown in eq.\eqref{NCparameter}. Let us now proceed to solve eq.\eqref{MassDimEq} for $m\neq0$. We do this by using the WKB approximation. We start with an ansatz of the following form
	\begin{eqnarray}
	\phi(r)\sim e^{S(r)}~.
	\end{eqnarray}
	This in turn gives us the following
	\begin{eqnarray}
	S^{\prime\prime}(r)+{S^{\prime}}^2(r)+\left(\frac{p(d+1)}{r}\right)S^{\prime}(r)-\frac{m^2}{r^{2p}}=0~.
	\end{eqnarray}
	Here, we focus on the leading order behaviour so we drop the term $S^{\prime\prime}(r)$ and solve it. This yields
	\begin{eqnarray}
	S^{\prime}(r)=\frac{-\frac{p(d+1)}{r}\pm\sqrt{\frac{4m^2}{r^{2p}}+\left(\frac{p(d+1)}{r}\right)^2}}{2}~.
	\end{eqnarray}
	In the non-conformal domain $p<1$. This in turn mean we can make the following approximation
	\begin{eqnarray}
	S^{\prime}(r)\approx -\frac{p(d+1)}{2r}\pm \frac{m}{r^p}~.
	\end{eqnarray}
	By intergrating the above, one finally obtains
	\begin{eqnarray}\label{Dimension}
	\phi(r)=r^{-\frac{p(d+1)}{2}}\exp{\pm\frac{m}{1-p}r^{1-p}}~.
	\end{eqnarray}
	The above result once again confirms the fact that there exists scaling symmetry as long as the mass $m$ is zero. However for $m\neq0$, it is not possible to derive a mass-conformal dimension. This is not unexpected as the background in hand holographically corresponds to a large $N$ QFT with irrelevant type non-conformal deformation. Therefore, in a strict sense, the concept of a conformal dimension of a dual operator loses its meaning in this context. In the subsequent analysis, we will only use the notion of conformal dimension whenever we consider the conformal limit, that is, $\eta\rightarrow0$.

\section{Pole-skipping points}\label{Sec2}
In recent times, it has become a well-known fact that there exist some special QNMs (in the complexified $\tilde{\omega}-\tilde{k}$ plane) at which the retarded Green's function of the concerned dual operator becomes ill-defined. Further, if the operator under consideration is the time-time component of the stress-tensor, then it is related to the parameters of chaos. In this section, we evaluate these points for the previously discussed (massive) scalar field perturbation in the bulk. Our aim is to analytically point out the location of the pole-skipping (PS) points in the complex $\tilde{\omega}-\tilde{k}$ plane and to check whether they are supported by the numerically computed spectral curves in the mentioned plane. This will also help us to classify the orders of the pole-skipping points. We start the computation by writing the NCBB in the ingoing Eddington-Finkelstein (EF) coordinates which are
\begin{eqnarray}
	r_*=\int\frac{dr}{r^{2p}f(r)};~v=t+r_*~.
\end{eqnarray}
Under the above coordinate transformations, the NCBB metric gets to the following form
\begin{eqnarray}
ds^2=-r^{2p}f(r)dv^2+2dv~dr+h(r)dx_idx^i;~h(r)=r^{2p}~.\nonumber\\
\end{eqnarray}
Recalling the KG equation (given in eq.\eqref{KGeq}) in the above background metric and considering the plane wave ansatz $\phi(r,v,x^i)=\int d^dk e^{-i\omega v+ik_ix^i}\psi_{\omega,k}(r)$ once again, leads us to
	\begin{eqnarray}\label{PSeq}
\mathcal{S}_{ps}:=\Delta_2(r)\psi_{\omega,k}^{\prime\prime}(r)+\Delta_1(r)\psi_{\omega,k}^{\prime}(r)+\Delta_0(r)\psi_{\omega,k}(r)=0\nonumber\\
	\end{eqnarray}
	where we have defined
	\begin{eqnarray}\label{eq0}
	\Delta_2(r)&=&h(r)^{\frac{d-1}{2}}r^{2p}f(r)\nonumber\\
	\Delta_1(r)&=&\partial_r\left[h(r)^{\frac{d-1}{2}} r^{2p} f(r)\right]-2i\omega h(r)^{\frac{d-1}{2}}\nonumber\\
	\Delta_0(r)&=&-i\omega\partial_r\left(h(r)^{\frac{d-1}{2}}\right)-h(r)^{\frac{d-3}{2}}\left(k^2+h(r)m^2\right)~.\nonumber\\
	\end{eqnarray}
	As a quick consistency check we have found that it produces the conformal result given in \cite{Blake:2019otz}, once we take the limit $\eta\rightarrow1$. We now expand eq.\eqref{PSeq} in the near-horizon region. For this we perform series expansion around $r=r_+$ for the quantities $\Delta_2(r),\Delta_1(r),\Delta_0(r)$ and assume the following near-horizon expansion for the wave-functional $\psi_{\omega,k}(r)=\sum_{k=0}^{\infty}\psi_{k}(r-r_+)^k$. This in turn leads us to
	\begin{eqnarray}\label{mainEQ}
	\mathcal{S}_{ps}=\mathcal{S}_{ps}^{(0)}+\mathcal{S}_{ps}^{(1)} (r-r_+)+\mathcal{S}_{ps}^{(2)}(r-r_+)^2+...=0~.\nonumber\\
	\end{eqnarray}
	Now by demanding that the above expanded form satisfies eq.\eqref{PSeq} in every order of $(r-r_+)$, we obtain an infinitely large set of coupled simple equations. This can be realised in the following way.\\
	In the leading order (LO) expansion, we have $\mathcal{S}_{ps}^{(0)}=0$ which has the following structure
	\begin{eqnarray}
	\Delta_2(r_+)\psi_{2}+\Delta_1(r_+)\psi_{1}+\Delta_0(r_+)\psi_{0}=0~.
	\end{eqnarray}
	From eq.\eqref{eq0}, we can see that $\Delta_2(r)$ on the horizon vanishes. Keeping this observation in mind and by evaluating $\Delta_0(r_+),\Delta_1(r_+)$ we obtain
	\begin{eqnarray}
	\mathcal{M}_{11}(\tilde{\omega},\tilde{k}^2)\psi_{0}+(i\tilde{\omega}-1)\psi_{1}=0
	\end{eqnarray}
where $\mathcal{M}_{11}(\tilde{\omega},\tilde{k}^2)= \left(\frac{1}{r_+}\right)\left[\frac{i}{2}p(d-1)\tilde{\omega}+\left(\frac{c}{4}\right)\tilde{k}^2+\frac{\tilde{m}^2}{c}\right]$~. Similarly, in the next-to leading order (NLO) in near-horizon expansion, one can show that
\begin{eqnarray}
\mathcal{M}_{21}(\tilde{\omega},\tilde{k}^2)\psi_{0}+\mathcal{M}_{22}(\tilde{\omega},\tilde{k}^2)\psi_{1}+(i\tilde{\omega}-2)\psi_{2}=0
\end{eqnarray}
where
\begin{widetext}
\begin{eqnarray}
\mathcal{M}_{22}&=&\left(\frac{1}{4r_+}\right)\left[\left(\frac{2}{c}\right)\left(c(c+1)-2pc(d+1)\right)+3i\tilde{\omega}p(d-1)+\left(\frac{c}{2}\right)\tilde{k}^2+\left(\frac{2}{c}\right)\tilde{m}^2\right]\nonumber\\
\mathcal{M}_{21}&=&\left(\frac{1}{4r_+^2}\right)\left[ip(d-1)\left(p(d-1)-1\right)\tilde{\omega}+p(d-3)\left(\frac{c}{2}\right)\tilde{k}^2+p(d-1)\left(\frac{2}{c}\right)\tilde{m}^2\right]~.
\end{eqnarray} 
\end{widetext}
This implies that from each order of eq.\eqref{mainEQ}, we obtain a linear equation which relates $\psi_0,\psi_{1},\psi_{2},...,\psi_n$ and collectively they a form a set of coupled algebraic equations. These coupled equations can be solved (upto the choice of the order) to obtain the pole-skipping points at which the two retarded Green's function $G^R_{\mathcal{O}_{\phi}\mathcal{O}_{\phi}}(\tilde{\omega},\tilde{k}^2)$ of the operator $\mathcal{O}_{\phi}$ which is dual to scalar field $\phi$ is ill-defined or more precisely, multi-valued. We now provide two of those set of points. These read as follows.
\begin{widetext}
$\mathrm{LO~expansion~and~PS~points:}$
	\begin{eqnarray}\label{PS1}
	\tilde{\omega}&=&-i\nonumber\\
	\tilde{k}&=&\pm\frac{i}{\sqrt{c}}\sqrt{\Big[2p(d-1)+\left(\frac{4}{c}\right)\tilde{m}^2\Big]}
	\end{eqnarray}
    $\mathrm{~~~NLO~expansion~and~PS~points:}$ 
\begin{eqnarray}\label{PS2}
\tilde{\omega}&=&-2i\nonumber\\
\tilde{k}&=&\pm i\sqrt{\frac{2}{c}}\sqrt{1 + c + \frac{2 \tilde{m}^2}{c} +p (d-3)\pm\sqrt{\mathcal{G}}}
\end{eqnarray}
where
\begin{eqnarray}
\mathcal{G}=(c+1)^2-2cp (d+1)+\frac{8 m^2 p}{c}+\left(d^2+10 d-7\right) p^2+(2-6 d) p\nonumber~.
\end{eqnarray}
One can follow these procedures to obtain higher order PS points in similar fashion.
\begin{figure}[!htb]
		\centering
		\begin{minipage}[t]{0.48\textwidth}
			\centering
			\includegraphics[width=\textwidth]{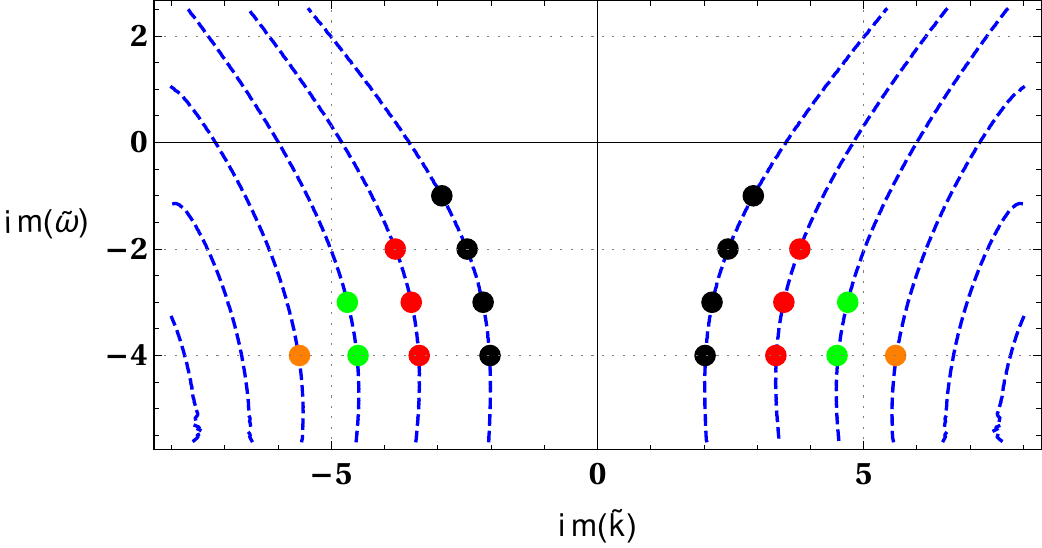}\\
			{\footnotesize  $~~~~~~~~~~~~~\eta=1$}
		\end{minipage}\hfill
		\begin{minipage}[t]{0.48\textwidth}
			\centering
			\includegraphics[width=\textwidth]{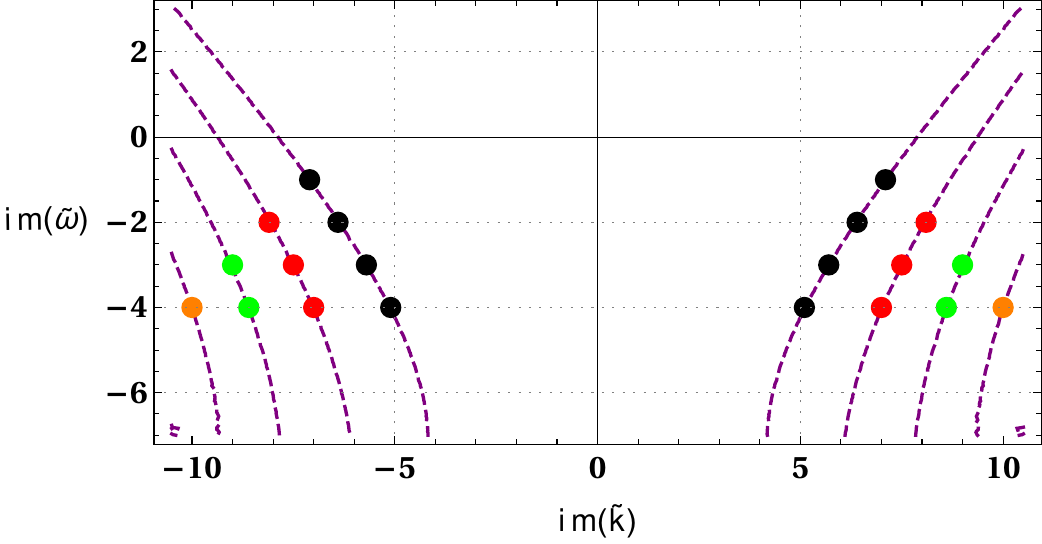}\\
			{\footnotesize  $~~~~~~~~~~~~\eta=2$}
		\end{minipage}
		\caption{Plot of the spectral curve of the massive scalar operator $\mathcal{O}_{\phi}$ for purely imaginary momenta and PS points obtained from upto fourth-order in near-horizon expansion. Here we consider $d=4$ and mass $m=2\sqrt{3}$.}
		\label{Plot1}
	\end{figure}
\end{widetext}
For the sake of better understanding about the order of the PS points, we now plot the spectral curve of the operator $\mathcal{O}_{\phi}$ for purely imaginary momenta (Re $\tilde{k}^2=0$) along with the obtained PS points. For $\eta=0,d=4$, that is for SAdS$_{4+1}$, it has been given in \cite{Abbasi:2020xli}. Here we focus on $\eta\neq0$ which corresponds to the non-conformal scenario. From Fig.\eqref{Plot1}, we observe that plot of the spectral curve of the massive scalar operator $\mathcal{O}_{\phi}$ for purely imaginary momenta (Re $\tilde{k}^2=0$) in turn gives the different dispersion relations satisfied by the QNMs of the massive scalar field $\phi$. It is also to be observed that if we plot the PS points in the same (im $\tilde{\omega}$-im $\tilde{k}$)-plane, it is possible to observe which set of PS points satisfy which QNM dispersion relation. For instance, one can observe that the PS points in black satisfy the same dispersion relation, on the other hand the dispersion relation satisfied by the red PS points are different. Keeping this in mind, we colour the PS points differently. From the above curves, it is also possible to see the effect of non-conformality on both the dispersion relation of the QNMs and the PS points.  
\begin{figure}[!h]
	\centering
	\includegraphics[width=0.44\textwidth]{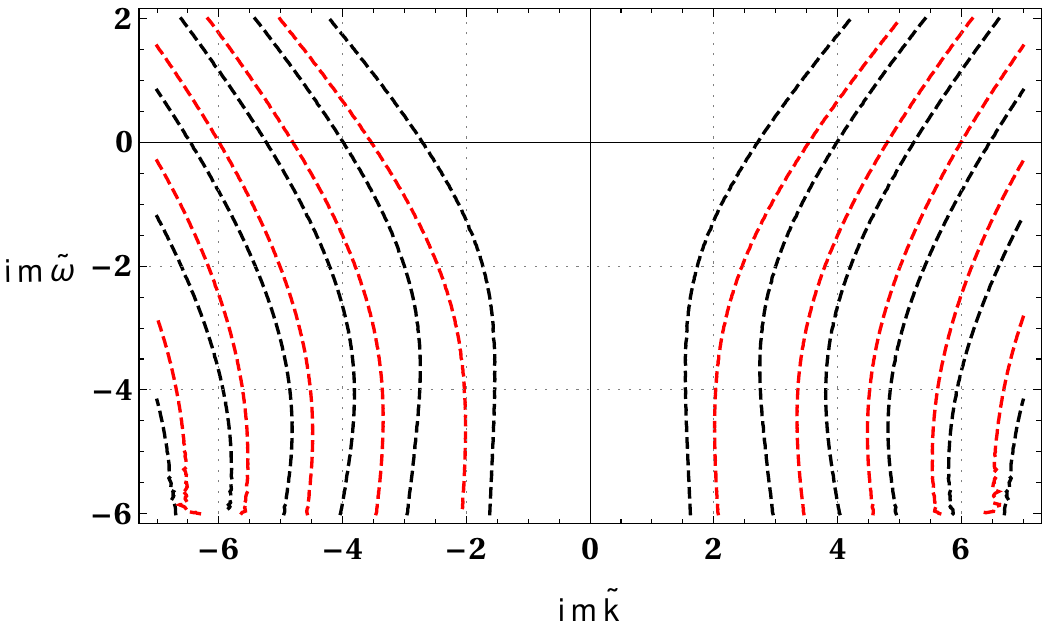}
	\caption{Effect of non-conformal parameter $\eta$ on the dispersion relations of QNMs where we set $d=4$ and $m=2\sqrt{3}$.}
	\label{Plot2}
\end{figure}
However, for the sake of better understanding, we compare them in a single plot which is given in Fig.\eqref{Plot2}. Here, the QNM dispersion relations given in black correspond to $\eta=0$ and same depicted in red correspond to $\eta=1$. The effect of non-conformality on the spectral curve of the massive scalar operator $\mathcal{O}_{\phi}$ can be noted affirmatively.
\section{Critical points of QNM spectral curve and convergence radii of derivative expansion}\label{Sec4}
In this section we intend to show the application of the QNM spectral curve in context of the derivative expansion of the gapped modes, keeping in mind the prescription given in \cite{Grozdanov:2019kge,Grozdanov:2019uhi}. 
\begin{widetext}
	As previously discussed, the scalar field fluctuations considered in this work yield QNMs that only satisfy a gapped dispersion relation. Further, we would like to stress the fact that this gapped nature is actually independent of the mass of the bulk scalar field. This can be understood from the plots given in Fig.\eqref{Plot3}. In this plots, we have numerically solved the spectral curve (by considering $40$ terms in the sum) to obtain the locations of the lowest four QNMs in the complex-$\tilde{\omega}$ plane. In this process we choose $\tilde{k}^2=0$ (zero momenta) and $\Delta_{\phi}=p(d+1)-1$ which in turn implies $\tilde{m}=0$ (one can also choose $\Delta_{\phi}=0$ to yield $\tilde{m}=0$ as we have shown in eq.\eqref{Dimension}). We observe that even the bulk field is massless, we do not have a QNM at $\tilde{\omega}=0$. This in turn means that the massless nature of the bulk field does not lead us to a gapless hydrodynamic mode, it is still gapped. This statement is also true for non-conformal scenario ($\eta\neq0$).
\begin{figure}[!htb]
	\centering
	\begin{minipage}[t]{0.42\textwidth}
		\centering
		\includegraphics[width=\textwidth]{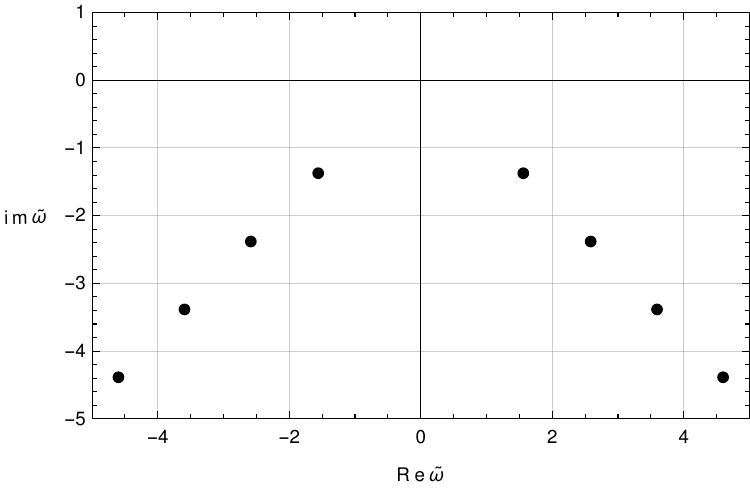}\\
		{\footnotesize  $~~~~~~~~~\eta=0$}
	\end{minipage}\hfill
	\begin{minipage}[t]{0.42\textwidth}
		\centering
		\includegraphics[width=\textwidth]{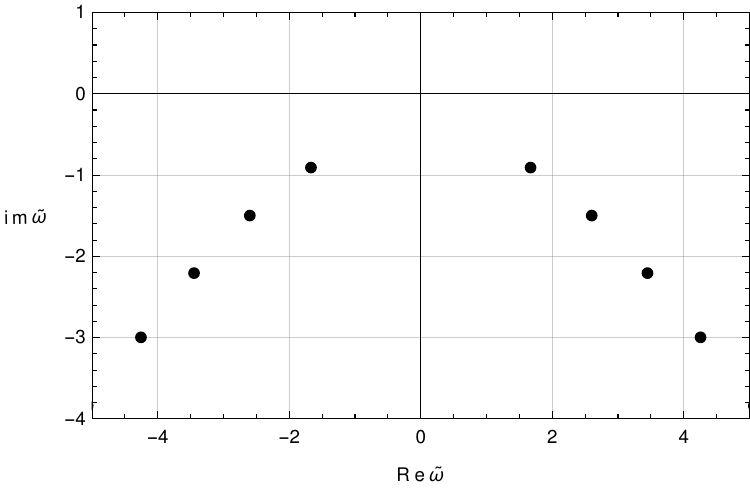}\\
		{\footnotesize  $~~~~~~~~~\eta=1$}
	\end{minipage}
	\caption{Lowest four QNMs for zero momenta and zero mass of the scalar field. We have set $d=4$.}
	\label{Plot3}
\end{figure}	
\end{widetext}
Here, our aim is to holographically find out the radius of convergence of the derivative expansion associated to this gapped mode dispersion relation. In simple sense, we wish to see whether the infinite series of the dispersion relation (which in fact captures the effect of all order in derivative expansion in the dual theory) converges or not. If yes, then the meaningful quantity to look for is the convergence radii. In \cite{Grozdanov:2019kge,Grozdanov:2019uhi}, it has been shown that the spectral curve $\mathcal{S}_{\phi}(\tilde{\omega},\tilde{k}^2,m)=0$ corresponding to the operator $\mathcal{O}_{\phi}$ is an implicit function which yields a spectral relation (or more physically, dispersion relation) in the $\mathbb{C}^2$ plane, as we have shown in eq.\eqref{DispFound} in the complex $(\tilde{\omega}-\tilde{k})$ plane. It is to be noted that here the coefficients of $\tilde{k}^2$ correspond to a particular branch of the Puiseux series expansion (sometimes also denoted as the generalized Taylor expansion).
\begin{widetext}
	 Now, keeping in mind the implicit function theorem and its properties one can compute the critical points which emerge from the solving of the following set of $\mathbb{C}$-equations
	\begin{eqnarray}\label{RadCon1}
	\mathcal{S}_{\phi}(\tilde{\omega},\tilde{k}^2,m)=0;~\partial_{\tilde{\omega}}\mathcal{S}_{\phi}(\tilde{\omega},\tilde{k}^2,m)=0~.	
	\end{eqnarray}
	\begin{figure}[!htb]
		\centering
		\begin{minipage}[t]{0.42\textwidth}
			\centering
			\includegraphics[width=\textwidth]{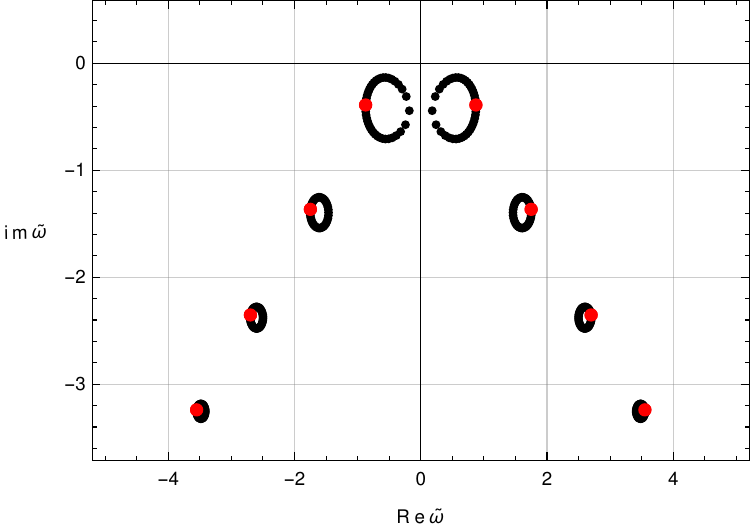}\\
			{\footnotesize  $~~~~~~~~|\tilde{k}|=0.55$}
		\end{minipage}\hfill
		%\par\vspace{4mm}
		\begin{minipage}[t]{0.42\textwidth}
			\centering
			\includegraphics[width=\textwidth]{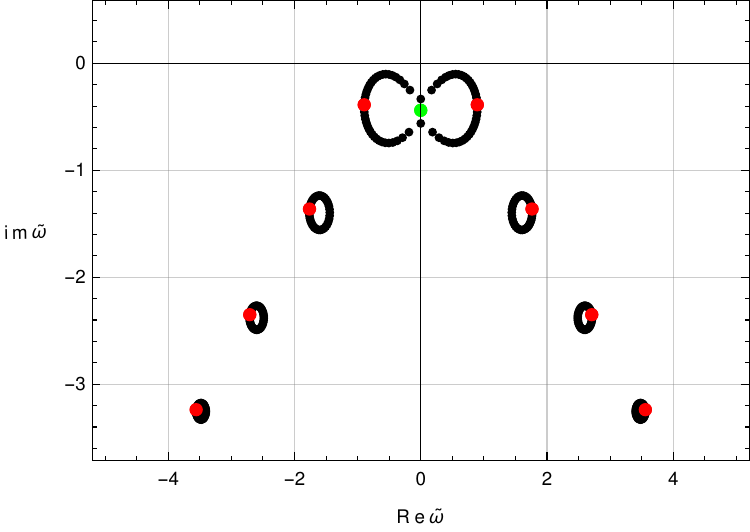}\\
			{\footnotesize  $~~~~~~~~|\tilde{k}|=0.57$}
		\end{minipage}\hfill
		\par\vspace{4mm}
		\begin{minipage}[t]{0.42\textwidth}
			\centering
			\includegraphics[width=\textwidth]{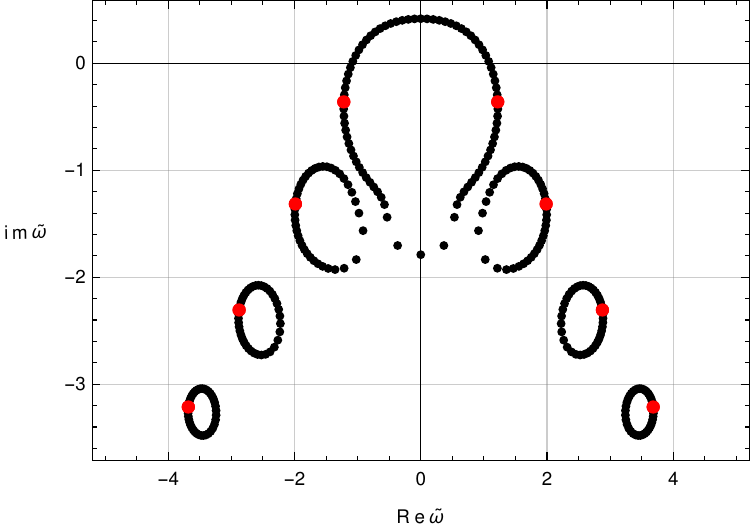}\\
			{\footnotesize  $~~~~~~~~|\tilde{k}|=0.97$}
		\end{minipage}\hfill
		\begin{minipage}[t]{0.42\textwidth}
			\centering
			\includegraphics[width=\textwidth]{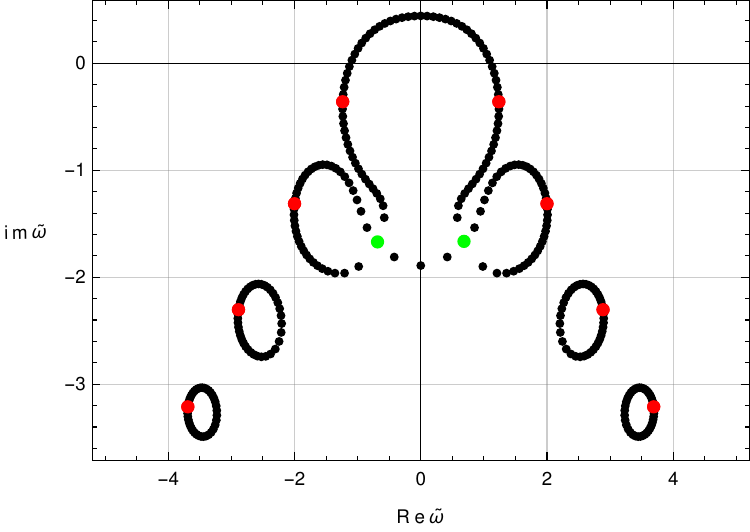}\\
			{\footnotesize  $~~~~~~~~|\tilde{k}|=0.98$}
		\end{minipage}\hfill
		\caption{We set $\Delta_{\phi}=2$ and plot the QNMs for $\tilde{k}^2=|\tilde{k}^2|e^{i\theta}$ with various values of $|\tilde{k}|$ and $0\leq\theta\leq 2\pi$. The left plot of the upper panel is for $|\tilde{k}|=0.55$, right plot of the upper panel is for $|\tilde{k}|=0.57$, left plot of the lower panel is for $|\tilde{k}|=0.97$ and the right plot in the lower panel is for $|\tilde{k}|=0.98$. Here, the red points depict the solutions for $\theta=0$ and green points represents the critical value of $\tilde{\omega}$ at which the QNMs collide.}
		\label{Plot4}
	\end{figure}
\end{widetext}
The radius of convergence is the fixed from that particular critical point $(\tilde{\omega}_c,\tilde{k}_c)$ which is closest to the origin, in our case $(0,\tilde{\omega}_0)$. To be precise, the convergence radii is $|\tilde{k}_c|$. We should also mention that $\tilde{\omega}_0$ have multiple solutions (we have shown this in Fig.\eqref{Plot3}) or more precisely, multiple orders of QNM. We will only focus on the solution with lowest (first order) and next-to lowest (second order) imaginary part as these two solutions are the longest living QNMs. This in turn mean the critical point closest to the first order solution of $\tilde{\omega}_0$ (at $\tilde{k}=0$) will give us the convergence radii of the lowest order QNM while the critical point closest to the second order solution of $\tilde{\omega}_0$ (at $\tilde{k}=0$) will lead us to the convergence radii of the second order QNM. On the other hand as we have already discussed that the spectral curve (implicit function) of the operator $\mathcal{O}_{\phi}$ is also the spectral curve for the QNMs for the dual bulk field $\phi$ and critical points can also be understood as the points at which two QNMs collide with each other. This in turn means that for given $\tilde{k}^2$ their real and imaginary part of the frequency are equal to each other. In order to evaluate these points one focuses on the value of $\tilde{\omega}(\tilde{k})$ in complex-$\tilde{\omega}$ plane for a given complex momentum in the following way $\tilde{k}^2=|\tilde{k}^2|e^{i\theta}$ where $\theta$ changes from $0$ to $2\pi$. This in turn means that from both these methods (collision of QNMs and implicit function property given in eq.\eqref{RadCon1}) can lead us to the desired radius of convergence. Some of the interesting studies in this direction can be found in \cite{Abbasi:2020ykq,Abbasi:2020xli,Asadi:2021hds,Taghinavaz:2023tog,Cartwright:2021qpp,Cartwright:2024rus,Cartwright:2024iwc}. We now proceed to compute the convergence radii for our with the motivation to quantify the effect of non-conformality on it. In the subsequent analysis we set $d=4$.\\
\begin{itemize}
	\item \textbf{For $\eta=0$:}\\
	
As we have mentioned already, for $\eta=0$ there is no deformation and the black brane solution is SAdS$_{4+1}$. This scenario has already been discussed in \cite{Abbasi:2020xli}. However, in order to understand the deviation of the obtained results from the conformal ones, we also discuss it here. In the conformal limit, one can make use of the mass-dimension relation given in eq.\eqref{MassDimStand} and express the mass of the bulk field in terms of the conformal dimension of the dual operator $\Delta_{\phi}$. We observe that as long as the dimension of the dual operator is less than a critical value $\Delta_{\phi}^c$, that is for $\Delta_{\phi}<\Delta_{\phi}^c$, the convergence radii of the first order QNM $\tilde{k}_c^{(1)}$ (lowest order solution of $\tilde{\omega}_0$) and same of the second order QNM $\tilde{k}_c^{(2)}$ are different. On the other hand, for $\Delta_{\phi}\geq\Delta_{\phi}^c$ it has been observed that these two convergence radii are same. 	From the QNM spectral curve, the nature of collisions (which leads to the convergence radii) are different as long as $\tilde{k}_c^{(1)}\neq\tilde{k}_c^{(2)}$. For the sake of better understanding we now switch to the graphical representations.
In Fig.\eqref{Plot4}, we present our observations for $\Delta_{\phi}=2$. Firstly, we should mention that by the term collision of the QNMs we mean that the closed trajectory of a mode (in $\theta$) becomes open upon colliding with the same for a different mode. For $|\tilde{k}|=0.57$, we observe that lowest order QNMS collide with each other along the im-$\tilde{\omega}$ axis. This type of collision has been dubbed as the lowest-level degeneracy. On the other hand, for $|\tilde{k}|=0.98$, we observe the collision between the first and second order QNMs. This has been dubbed as the level-crossing of QNMs. This in turn means that the convergence radii about the first QNM is $|\tilde{k}_c|=0.57$ and $|\tilde{k}_c|=0.98$ denotes the convergence radii about the second order QNM. We have found that in the $\eta=0$ scenario, as long as $\Delta_{\phi}<5.2$, one can observe that the convergence radii corresponding to first order and second order QNMs are different. We denote this particular value of the operator dimension as $\Delta_{\phi}^c$ which in this case is $\Delta_{\phi}^c\approx 5.2$ which corresponds to critical mass $m_c\approx 2.54$. At this particular value of $\Delta_{\phi}$, lowest-level degeneracy (between the first order QNMs) and level-crossing s (between first and second QNMs) simultaneously occur. On the other hand, beyond this value of dimension, convergence radii for the first and second order QNMs same as only level-crossing type collision occurs.\\

\begin{figure}[!h]
	\centering
	\includegraphics[width=0.4\textwidth]{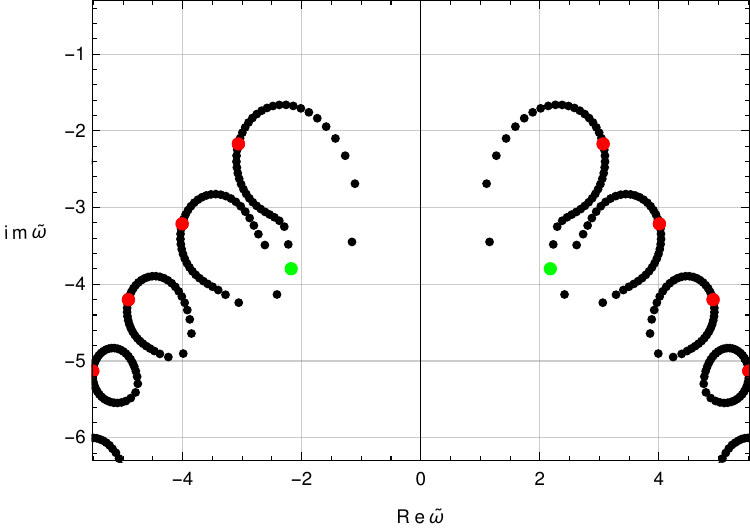}
	\caption{Level-crossing type collision between first and second order QNMs at $\Delta_{\phi}=6$}
	\label{Plot5}
\end{figure}

This we represent in Fig.\eqref{Plot5} where we have considered $\Delta_{\phi}=6$. We observe that only level-crossing type collision is occurring. We now give a table for our obtained result for convergence radii about the first and second order QNMs corresponding to few values of $\Delta_{\phi}$.
\begin{table}[h]
	\begin{tabular}{ ||c|c|c|c|| } 
		\hline
		$\Delta_{\phi}$ & $m$ & value of $\tilde{\omega}_c$ & convergence radii $|\tilde{k}_c|$  \\
		\hline 
		2 & 2i &-0.44i & 0.57 \\ 
		3 & $\sqrt{3}$i & -0.97i & 0.87 \\
		4 & 0 &-1.81i & 1.13 \\ 
		5 & $\sqrt{5}$ &-2.65i & 1.34 \\ 
		6 & $2\sqrt{3}$ &$\pm2.17-3.79$i  & 1.48 \\ 
		\hline
	\end{tabular}
	\caption{Values of radius of convergence about the first order QNM and corresponding values of $\tilde{\omega}_c$}
	\label{Table1}
	\end{table}
	\begin{table}[h]
		\begin{tabular}{ ||c|c|c|c|| } 
			\hline
			$\Delta_{\phi}$ & $m$ &value of $\tilde{\omega}_c$ & convergence radii $|\tilde{k}_c|$  \\
			\hline 
			2 & 2i &$\pm0.68-1.67$i & 0.97 \\ 
			3 & $\sqrt{3}$i &$\pm1.18-1.91$i & 1.05 \\
			4 & 0 & $\pm1.54-2.76$i & 1.25 \\ 
			5 & $\sqrt{5}$ & $\pm1.74-3.23$i &  1.36 \\ 
			6 & $2\sqrt{3}$ & $\pm2.17-3.79$i  & 1.48 \\ 
			\hline
		\end{tabular}
		\caption{Values of radius of convergence about the second order QNM and corresponding values of $\tilde{\omega}_c$}
		\label{Table2}
	\end{table}
	In Table \eqref{Table1} and Table \eqref{Table2}, we give the numerical values of convergent radii with corresponding $\tilde{\omega}_c$ (at which they collide) for first order and second order QNM. We have also mentioned the mass values corresponding to the values for $\Delta_{\phi}$. Our observations made for the $\eta=0$ scenario, qualitatively matches with the one given in \cite{Abbasi:2020xli}. 
	\item \textbf{For $\eta\neq0$:}\\

	When we switch on the non-conformal deformation by setting a particular value for the deformation parameter $\eta$ (say $\eta=1$ or $2$), the effects can be observed prominently. As we have already shown that in the presence of the non-conformal deformation ($\eta\neq0$), it is not possible to derive a definite a mass-dimension relation and there is no notion of conformal dimension. Keeping in mind these observations, for $\eta\neq0$ we just focus on the mass of the bulk field $m$. Furthermore, as the mass-dimension relation is not available for $\eta\neq0$, the corresponding Breitenlohner-Freedman (BF) bound is also not known. Due to this, we restrict ourselves only to $m\geq0$. Firstly, we observe that similar to the $\eta=0$ case, there exist a particular value of $\Delta_{\phi}^c$ (or mass value $m_c$) upto which both lowest-level degenerate type and level-crossing type collision occurs. We found that for $\eta=1$, this critical mass value is $m_c\approx2.2$. We now choose two different values for the mass $m$, one in the range $m<m_c$ and one in the range $m>m_c$ in order to observe this. In Fig.\eqref{Plot6}, we observe that for $m=2$ the lowest-level degenerate type collision occurs for $|\tilde{k}|=1.66$ whereas the level-crossing happens at $|\tilde{k}|=1.72$. These plots also imply that for a given value of the bulk field mass $m$, non-conformality increases the value of convergence radii.
	\begin{widetext}
		~~\\
		\begin{figure}[!htb]
			\centering
			\begin{minipage}[t]{0.42\textwidth}
				\centering
				\includegraphics[width=\textwidth]{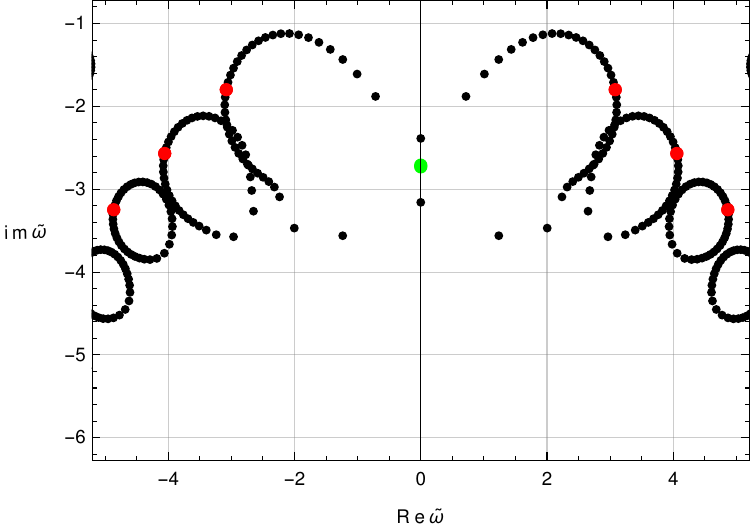}\\
			\end{minipage}\hfill
			\begin{minipage}[t]{0.42\textwidth}
				\centering
				\includegraphics[width=\textwidth]{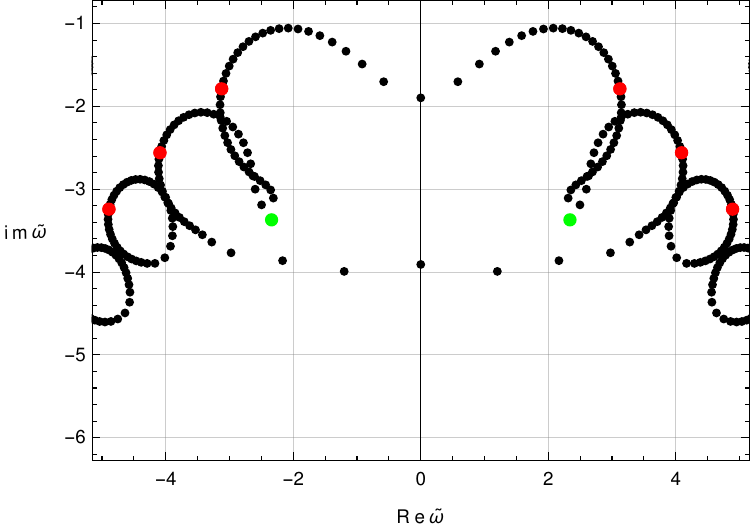}\\
			\end{minipage}
			\caption{In these plots we set $m=2$ and $\eta=1$. The upper plot represents lowest-level degeneracy corresponding to the value $|\tilde{k}|=1.66$ and the lower plot represents level-crossing corresponding to the value $|\tilde{k}|=1.72$.}
			\label{Plot6}
		\end{figure}\\
Once again similar to the $\eta=0$ case, we choose $m=3$ (which is beyond $m_c$). We observe that only level-crossing occurs and it happens for $|\tilde{k}|=1.95$. In Fig.\eqref{Plot7} we provide this observation.
	\begin{figure}[!htb]
		\centering
		\begin{minipage}[t]{0.42\textwidth}
			\centering
			\includegraphics[width=\textwidth]{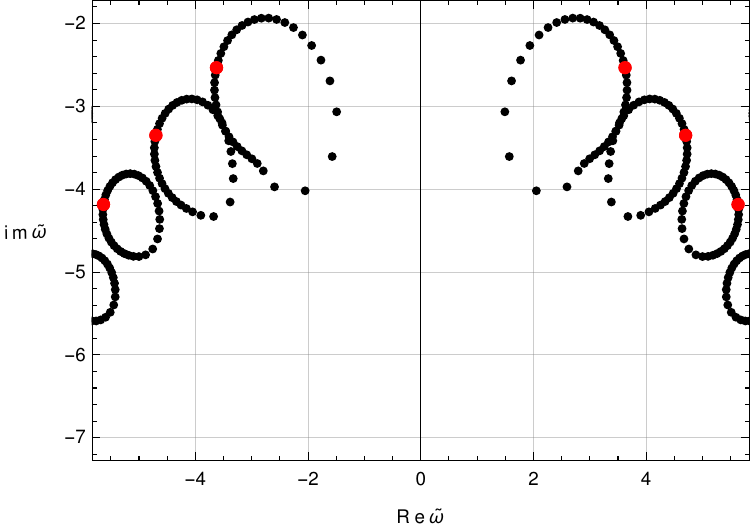}\\
		\end{minipage}\hfill
		\begin{minipage}[t]{0.42\textwidth}
			\centering
			\includegraphics[width=\textwidth]{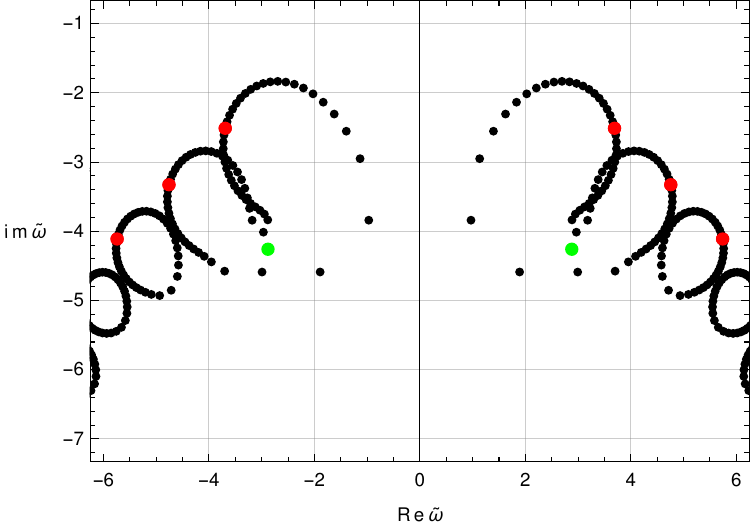}\\
		\end{minipage}
		\caption{We have set $m=3$ and $\eta=1$. The left plot corresponds to the value $|\tilde{k}|=1.8$ and the right plot corresponds to the value $|\tilde{k}|=1.82$.}
		\label{Plot7}
	\end{figure}
	\begin{table}[h]
		\begin{tabular}{ ||c|c|c|| } 
			\hline
			value of $m$ & value of $\tilde{\omega}_c$ & convergence radii $|\tilde{k}_c|$  \\
			\hline 
			0 & -1.24i & 1.23 \\ 
			1 & -1.76i & 1.45 \\
			2 & -2.71i & 1.66 \\ 
			3 & $\pm2.88-4.26$i & 1.82 \\ 
			4 & $\pm3.42-5.27$i  & 1.99 \\ 
			\hline
		\end{tabular}
		\caption{Values of radius of convergence about the first order QNM and corresponding values of $\tilde{\omega}_c$}
		\label{Table3}
		\begin{tabular}{ ||c|c|c|| } 
			\hline
			value of $m$ & value of $\tilde{\omega}_c$ & convergence radii $|\tilde{k}_c|$  \\
			\hline 
			0 & $\pm1.56-2.17$i & 1.45 \\ 
			1 & $\pm1.99-2.38$i & 1.58 \\
			2 & $\pm2.34-3.351$i & 1.71 \\ 
			3 & $\pm2.88-4.26$i & 1.82 \\ 
			4 & $\pm3.42-5.27$i  & 1.99 \\ 
			\hline
		\end{tabular}
		\caption{Values of radius of convergence about the second order QNM and corresponding values of $\tilde{\omega}_c$}
		\label{Table4}
	\end{table}\\
\end{widetext}
We now provide the radii of convergences for various values of $m$ and compare them with the same for $\eta=0$ scenario. In Table \eqref{Table3} and Table \eqref{Table4}, we provide the numerical values we have obtained for various values of the bulk field mass for $\eta=1$. By comparing them with the same computed for $\eta=0$, we can conclude that presence of non-conformal deformation enhances the radius of convergence of the derivative expansion. This in turn means that the presence of non-conformality increases the domain of validity (radius of convergence) for the derivative expansion which is a perturbative approach to the theory of hydrodynamics.\\
We further increase the value of $\eta$ to $\eta=2$ and repeat the same computations once again. Interestingly, for $\eta=2$ we observe that only level-crossing type of collision survives for the dimension range $0\leq\Delta_{\phi}\leq4$. This in turn means for a sufficiently large value of $\eta$, the convergence radii for both first and second order QNMs are same. The rest of the observations are qualitatively same, that is, the value corresponding to radius of convergence increases further. This we provide in Table \eqref{Table5}.
\begin{table}[h]
	\begin{tabular}{ ||c|c|c|| } 
		\hline
		value of $m$ & value of $\tilde{\omega}_c$ & convergence radii $|\tilde{k}_c|$  \\
		\hline 
		0 & $\pm 1.79-2.25$i & 1.52 \\ 
		1 & $\pm 2.55-2.34$i & 1.69 \\
		2 & $\pm 3.21-3.60$i & 1.85 \\ 
		3 & $\pm 5.02-6.73$i & 2.05 \\ 
		4 & $\pm 5.57-9.24$i & 2.38 \\ 
		\hline
	\end{tabular}
	\caption{Values of radius of convergence for both first and second order QNMs and corresponding values of $\tilde{\omega}_c$}
	\label{Table5}
	\end{table}
\end{itemize}
It is to be mentioned that the deformation parameter is bounded from the above as $\eta< \sqrt{\frac{8d}{d-1}}$ which for $d=4$ boils down to $\eta<3.26$. This Gubser bound implies any value of $\eta$ greater than this leads to a completely thermodynamically unstable black brane solution. Keeping this in mind, we restrict ourselves to $\eta=1,2$. Further, our aim is to see the effect of non-deformation on the convergence radii for the gapped modes which we have already succeeded to do above. 
\section{Comparison with the pole-skipping points and some insights}\label{Sec5}
In this section we compare the absolute value of the momenta corresponding to the pole-skipping point ($|\tilde{k}_*|$) which is closest to the origin to the obtained results of convergence radii in the momentum space ($|\tilde{k}_c|$). 
\begin{figure}[h]
	\centering
	\includegraphics[width=0.4\textwidth]{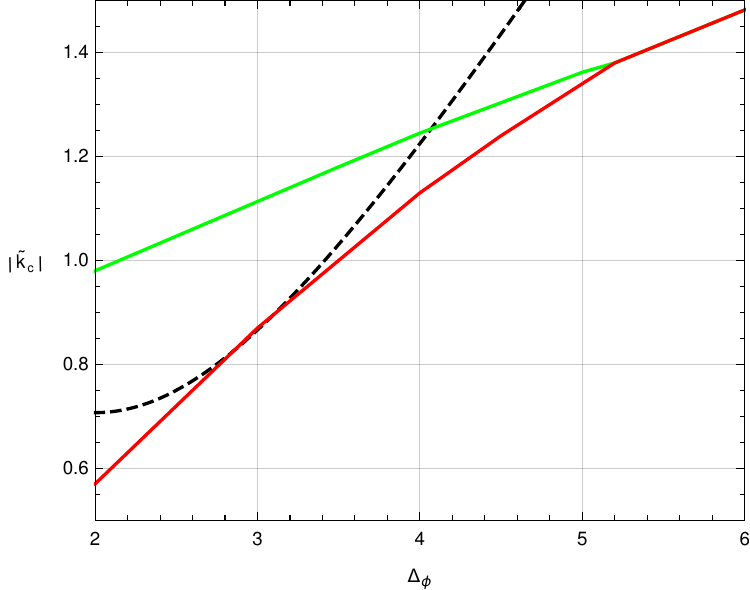}
	\caption{Comparison between $\tilde{k}_*$ (in dashed) and $\tilde{k}_c$ associated to both first order QNM (in red), second order QNM (in green). In this case $\eta=0$.}
	\label{Plot8}
\end{figure} 
\begin{figure}[!htb]
		\centering
		\begin{minipage}[t]{0.42\textwidth}
			\centering
			\includegraphics[width=\textwidth]{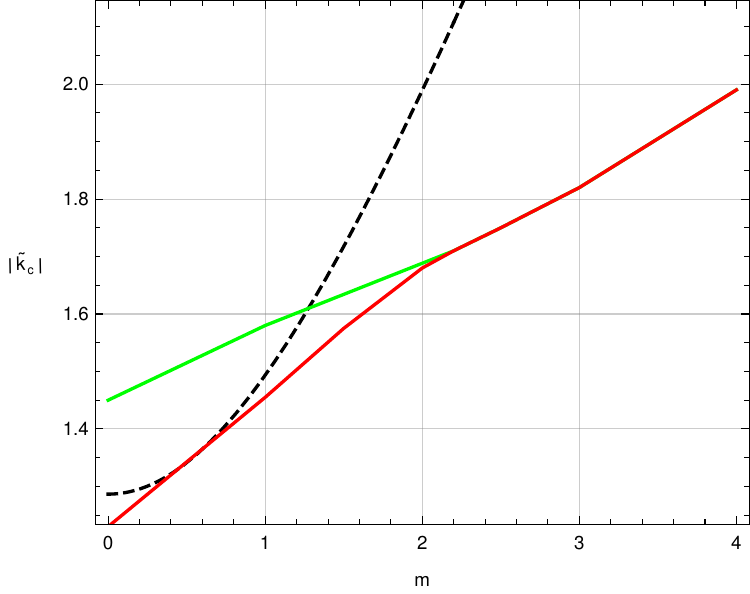}\\
			{\footnotesize  $\eta=1$}
		\end{minipage}\hfill
		\begin{minipage}[t]{0.42\textwidth}
			\centering
			\includegraphics[width=\textwidth]{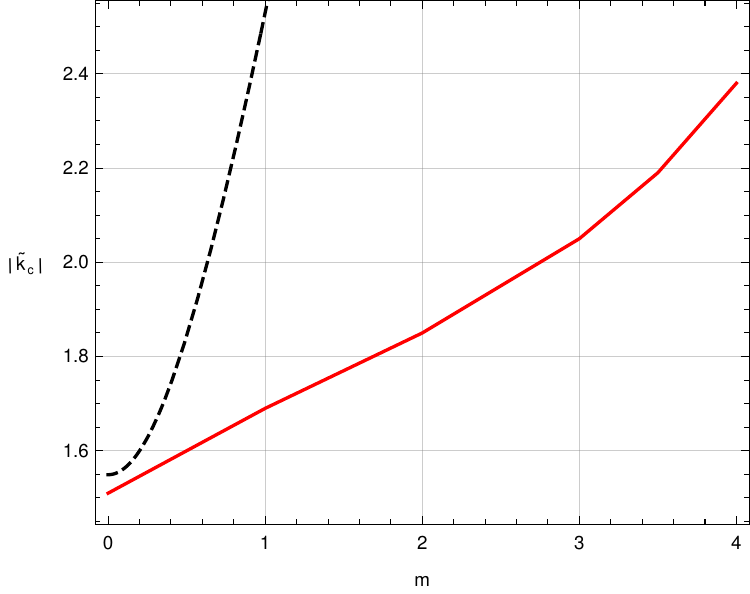}\\
			{\footnotesize  $\eta=2$}
		\end{minipage}
		\caption{Comparison between $\tilde{k}_*$ (in dashed) and $\tilde{k}_c$ associated to both first order QNM (in red), second order QNM (in green). Upper plot corresponds to $\eta=1$ and the lower plot corresponds to $\eta=2$.}
		\label{Plot9}
	\end{figure}
The reason behind this is the following. The derivative expansion is a perturbative approach associated to the hydrodynamic effective description of a system. This in turn means that convergence radii corresponds to the long wavelength description or IR representation of the system under consideration. On the other hand, pole-skipping phenomena is associated with exact structure of the retarded Green's function (for a relevant operator) of the theory. This implies pole-skipping represents the UV picture of the theory as the origin is purely quantum mechanical. Comparison between $|\tilde{k}_*|$ and $|\tilde{k}_c|$ can lead us to two things. If $|\tilde{k}_c|<|\tilde{k}_*|$, then one can say that with the help of a perturbative approach (such as derivative expansion) it is possible to probe into the UV picture of the theory as one is able to observe pole-skipping and if $|\tilde{k}_c|>|\tilde{k}_*|$ then one needs to follow a non-perturbative approach for hydrodynamics in order to be able to look into the UV picture of the theory, in short, a quantum theory of hydrodynamics is needed \cite{Blake:2017ris}. To proceed further, we recall the result for the mentioned pole-skipping point from eq.\eqref{PS1} and obtain the corresponding absolute value of momenta. This reads
\begin{eqnarray}
|\tilde{k}_*|=\frac{1}{\sqrt{c}}\sqrt{\Big[2p(d-1)+\left(\frac{4}{c}\right)\tilde{m}^2\Big]}~.
\end{eqnarray}
We now compare the above value of $|\tilde{k}_*|$ with the obtained values for $|\tilde{k}_c|$ at a given $\Delta_{\phi}$. In Fig.\eqref{Plot8}, we do this for $\eta=0$. It can be observed that the convergence radii associated to the first order QNM (in red) always bounded from the above by $|\tilde{k}_*|$. However, for a range of $\Delta_{\phi}$, the convergence radii of the second order is smaller than $|\tilde{k}_*|$. This implies from the perspective of the first order QNM, derivative expansion is not enough to observe the UV picture for the theory. One has implement a non-perturbative approach. Although, for a range of $\Delta_{\phi}$, the convergence radii of the second order QNM is smaller than $|\tilde{k}_*|$ but this observation does not yield something hopeful. The reason is the following. We need to keep in mind that here we are considering the pole-skipping point closest to the origin and it shares the same dispersion relation only with the first order QNM. So, from the closest to the origin pole-skipping point, only the convergence radii corresponding to first order QNM is meaningful. In Fig.\eqref{Plot9}, we observe that same observation is also true for the $\eta\neq0$ scenario. Furthermore, from these plot we can also observe that for $\eta=1$ one can write down the following for the convergence radii for the first order QNM 
\begin{eqnarray}
|\tilde{k}_*|\geq |\tilde{k}_c|~.
\end{eqnarray}
This is due to the fact for a particular value of $m$ this bound saturates. However, for $\eta=2$, this bound never saturates. This in turn means presence of non-conformal deformation advocates for a non-perturbative approach for hydrodynamics which makes one able to probe into the UV picture.
\section{Conclusion}\label{Sec6}
We now summarize our finding to conclude. In this work, we consider an irrelevant type of non-conformal deformation of the gauge theory in the holographic set up. This has been done by considering Einstein-dilaton theory with Liouville type dilaton potential as the bulk theory which has a non-AdS, warped asymptotic structure. This bulk theory has a finite temperature, black brane solution which we denote as the non-conformal black brane geometry. We then perturb this geometry by introducing a minimally-coupled massive scalar field $\phi$. By following a Frobenius type near-horizon expansion, we obtain the spectral curve of quasi-normal modes for this massive scalar field. By keeping in mind the essence of gauge/gravity duality, we holographically denote this quantity as the spectral curve for the scalar operator $\mathcal{O}_{\phi}$ which is nothing but the operator dual of the field $\phi$. By considering purely imaginary momenta, we show the various orders of QNMs (poles of the two-point retarded Green's function of the scalar operator $\mathcal{O}_{\phi}$) which are characterized by different dispersion relations. We also highlight the fact that these dispersion relations are gapped in nature. The location of the pole-skipping points associated to the relevant Green's function are also pointed out in this plane. This also helps us to see which set of pole-skipping points satisfy which dispersion relation. The effects of non-conformality on the spectral curve and the pole-skipping points can be observed prominently. Next we proceed to compute the convergence radii (in momentum space) of the derivative expansion for the gapped QNMs, denoted as $|\tilde{k}_c|$. This we have done by computing the critical points of the spectral curve $\mathcal{S}_{\phi}(\tilde{\omega},\tilde{k}^2,m)$ which is interestingly an implicit function. It has been observed that upto a critical value of the operator dimension $\Delta_{\phi}^c$ (or critical value for the mass $m_c$), the convergence radii for the first order QNM is different from the same obtained for second order QNM and they are characterized by two different type of collisions, namely, lowest-level degenerate type and level-crossing type respectively. We found that presence of non-conformal deformation increases the convergence radii of the QNMs for a given mass of the bulk field $m$. This in turn means that non-conformality enhances the domain of applicability for the derivative expansion. Finally, by comparing $|\tilde{k}_c|$ with the absolute momenta of lowest order pole-skipping point $|\tilde{k}_*|$, we realize that perturbative approach of hydrodynamics such as derivative expansion is not enough to probe the ultra-violet sector of the theory. This realization comes from the fact that convergence radii is always bounded from the above by $|\tilde{k}_*|$.
\bibliographystyle{hephys}   
\bibliography{Reference}

\end{document}